\documentclass[aps,prl,reprint,superscriptaddress,floatfix,footinbib,showkeys]{revtex4-1}
\usepackage[utf8]{inputenc}
\usepackage{amsfonts,amssymb,amsmath}
\usepackage{hyperref}
\usepackage{graphicx}

\newcommand{\mr}[1]{\mathrm{#1}}
\newcommand{\be}{\begin{equation}}
\newcommand{\ee}{\end{equation}}

\newcommand{\kb}{k_{\mr{B}}}

\newcommand{\sis}{\mr{S_1IS_2}}
\newcommand{\rt}{R_\mr{T}}
\newcommand{\vsis}{V_\mathrm{SIS}}
\newcommand{\nqp}{n_\mathrm{qp}}
\newcommand{\tb}{T_\mathrm{b}}
\newcommand{\vb}{V_\mathrm{b}}
\newcommand{\nng}{n_\mathrm{g}}
\newcommand{\vg}{V_\mathrm{g}}
\newcommand{\cg}{C_\mathrm{g}}
\newcommand{\dfermi}{D\left(E_\mathrm{F}\right)}
\newcommand{\ec}{E_\mathrm{c}}
\newcommand{\ach}{A_\mathrm{ch}}
\newcommand{\pinj}{P_\mathrm{inj}}

\begin{document}
\title{Active quasiparticle suppression in a non-equilibrium superconductor}
\author{Marco Marín-Suárez}\email{marco.marinsuarez@aalto.fi}
\author{Joonas T. Peltonen}
\author{Jukka P. Pekola}
\affiliation{Pico group, QTF Centre of Excellence, Department of Applied Physics, Aalto University, FI-000 76 Aalto, Finland}

\begin{abstract}
Quasiparticle (qp) poisoning is a major issue that impairs the operation of various superconducting devices. Even though these devices are often operated at temperatures well below the critical point where the number density of excitations is expected to be exponentially suppressed, their bare operation and stray microwave radiation excite the non-equilibrium qps. Here we use voltage-biased superconducting junctions to  demonstrate and quantify qp extraction in the turnstile operation of a superconductor-insulator-normal metal-insulator-superconductor single-electron transistor. In this operation regime excitations are injected into the superconducting leads at a rate proportional to the driving frequency. We reach a reduction of density by an order of magnitude even for the highest injection rate of $2.4\times 10^8$ qps per second when extraction is turned on.
\end{abstract}


\maketitle

In superconducting circuits it is important to minimize the number of non-equilibrium quasiparticles as they deteriorate the operation of various devices, such as the coherence of quantum bits based on Josephson junctions~\cite{Martinis2009,Paik2011,Barends2011,Catelani2011,Lenander2011} or Majorana nanowires~\cite{Rainis2012,Higginbotham2015}, cooling power of superconducting microcoolers~\cite{Muhonen2012,Nguyen2013}, sensitivity of kinetic inductance detectors~\cite{Visser2011,Visser2012,Patel2017} and performance of superconducting resonators in other applications~\cite{Nsanzineza2014,Gruenhaupt2018}. In principle bringing the system to temperatures $T$ much below the superconducting transition should reduce the number of excitations, as at $\kb T\ll\Delta$ (here $\kb$ is the Boltzmann constant and $\Delta$ the superconducting energy gap) their equilibrium number density $\nqp$ is suppressed exponentially. It has been demonstrated, however, that when the devices are operated, quasiparticle (qp) excitations are created, caused typically by the drive signals or by stray microwave photons from hotter stages of the refrigerator~\cite{Martinis2009,Knowles2012,Maisi2013,Gustavsson2016,Saira2010}, or by ionizing radiation~\cite{Vepsalainen2020}. To overcome this ``qp poisoning", several methods have been studied. These include introduction of normal metal traps~\cite{Goldie1990,Riwar2016,Hosseinkhani2018}, geometry optimization~\cite{Knowles2012,Wang2014}, vortex traps by magnetic field~\cite{Wang2014,Nsanzineza2014,Vool2014,Taupin2016}, gap engineering by variation of the film thickness~\cite{Aumentado2004,Yamamoto2006,Sun2012,Riwar2019} or phonon traps~\cite{Henriques2019,Karatsu2019}. Recently, blockage of qps by a voltage filter-tuned superconducting gap has been demonstrated~\cite{Menard2019}, and employing a lower gap superconductor as a qp trap has been analyzed in detail~\cite{Riwar2019}. Although the crucial role of low qp density was identified already in early studies of superconducting qubits~\cite{Lang2003,Lutchyn2005,Shaw2008} and related single-charge circuits~\cite{Aumentado2004,Maennik2004}, further understanding of the generation mechanisms and reducing $\nqp$ remains the topic of an ever increasing intense research activity~\cite{Riste2013,Pop2014,Gustavsson2016,Serniak2018,Serniak2019} as the effort to increase the coherence times of superconducting qubits continues.

\begin{figure}[ht!]
	\includegraphics[scale=0.295]{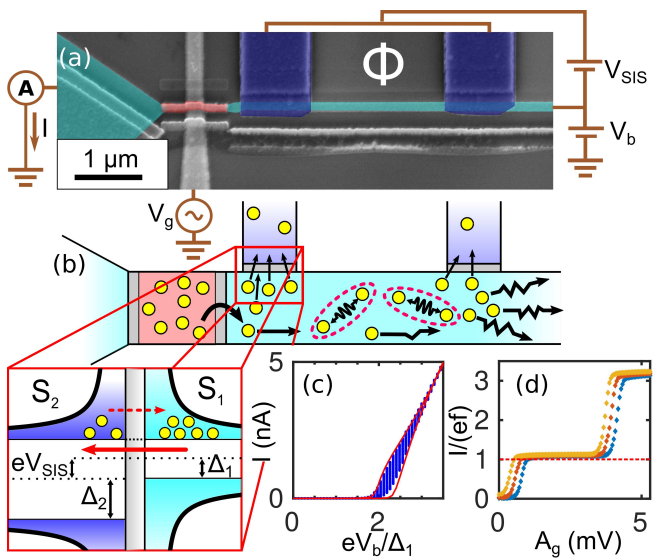}
	\caption{Qp extraction probed in a SINIS SET. (a) Electronic scheme for turnstile operation of the hybrid SET with an integrated $\sis$ cooler, together with a false color scanning electron microscope image of sample A. Light red marks the normal metal and blue the superconducting parts, with light (dark) blue designating the lower (higher) gap superconductor. (b) Schematic sketch of the device. Quasiparticles are injected to the narrow lead where relaxation is inefficient. Qp excitations are extracted via biased $\sis$ junctions, and recombination and scattering processes are still present along the lead. The inset shows that when the $\sis$ junction is biased at $e\vsis=|\Delta_1-\Delta_2|$ the singularities in the qp densities of states match and excitations are transferred from the lead with $\Delta_1$ to the reservoir with $\Delta_2$. (c) IV curves of sample B in the positive bias regime, for a large number of different gate voltages. The red lines are simulations for the envelope curves, and the blue dots are the experimental data. (d) Measurements for sample A in the turnstile operation at $f=10\,\mr{MHz}$, $\vb=120,\,160$ and $200\,\mr{\mu V}$ as blue, red and yellow symbols, respectively. Here the current is measured against the amplitude of the gate signal $A_\mr{g}$ and the red dashed line illustrates the ideal value of $I=ef$ for the first plateau.}
	\label{f1}
\end{figure}

For radio frequency (rf) driven superconductor-insulator-normal metal-insulator-superconductor (SINIS)\- turnstiles~\cite{Pekola2007}, operated as a source of quantized electric current~\cite{Pekola2013}, both drive-induced and background qps are the most severe limitation to reaching metrological current quantization accuracy~\cite{Knowles2012,Pekola2013}. In this work we show that such a hybrid single-electron transistor (SET, see Fig.~\ref{f1}(a) for a typical sample) functions as a sensitive, practical, and quantitative detector of the qp density of its superconducting electrodes, that could be integrated with a variety of other mesoscopic superconducting devices to probe their $\nqp$. Under rf drive, hybrid SETs present a turnstile for single electrons, a simple-to-operate candidate realization for a solid-state standard of electric current~\cite{Pekola2007,Maisi2009}. For a wide range of parameters, the output current $I=ef$ is determined only by the drive frequency $f$ and electron charge $e$ (see Fig.~\ref{f1}(d) for typical measurements in this regime). Here the non-equilibrium state results from qp injection (Fig.~\ref{f1}(b)) when the drive signal at frequency $f$ is applied to the gate electrode of the transistor. Earlier work~\cite{Knowles2012} demonstrated that the influence of drive-induced and environmental qps to a SINIS turnstile can be reduced by improving the geometry of the S electrodes and by shielding from residual microwave radiation, respectively.

Here we combine the turnstile with an independent, \emph{in-situ} control -- both extraction and injection -- of qps. The qp poisoning can be reduced by suitably voltage biasing superconductor-insulator-superconductor ($\sis$) junctions with different superconducting gaps ($\Delta_1$ and $\Delta_2$, respectively) where excitations are extracted from $\mr{S_1}$ to $\mr{S_2}$ as long as $\Delta_1<\Delta_2$~\cite{Chi1979,Blamire1991,Heslinga1993,Manninen1999}. In an important experiment, this effect has been used to cool one of the leads of a single-Cooper pair transistor~\cite{Ferguson2008} in a similar manner as refrigeration by normal metal-insulator-superconductor junctions~\cite{Pekola2000,Giazotto2006,Rajauria2009} has been used to cool down the normal lead. However, when using a single Cooper-pair transistor it was not possible to control the mechanism or rate of qp creation, making it more difficult to quantify the population reduction due to the biased $\sis$ junction. On the contrary, hybrid single-electron transistors allow for quantitative control. To that end, here we demonstrate that direct $\sis$ \emph{cooling} of the turnstile leads offers promise to fully extract the drive-induced qps under typical pumping conditions, in particular when bulky or thick electrodes cannot be utilized.

The evacuation of qps is manifested by the stabilization of current to $ef$ in the SET turnstile operation~\cite{Knowles2012}. The qp extraction can be tuned by varying the voltage biasing of the $\sis$ junction. However, non-equilibrium qps are also subjected to recombination and diffusion processes along the electrode as depicted in panel (b) of Fig.~\ref{f1}. Due to the exponential dependence of $\nqp\approx\dfermi\sqrt{2\pi\Delta\kb T}e^{-\Delta/\kb T}$ on $\Delta$ the number of excitations is lower in the higher gap film. Here $\dfermi$ denotes the normal-state density of states at the Fermi energy and $T$ is the electronic temperature of the superconductor. Therefore, providing sufficient energy to qps in the lower gap superconductor, biasing the junction such that the singularities in the superconducting densities of states align, will promote a transfer to the higher-gap superconductor (see the inset of Fig.~\ref{f1}(b))~\cite{Parmenter1961}. Although two superconductors with different energy gaps are required, the qp extraction in a voltage-biased tunnel junction is in sharp contrast to those gap engineering methods~\cite{Aumentado2004,Yamamoto2006,Sun2012,Riwar2019} where qps are passively trapped in lower-gap regions away from the relevant operational zones of the device. To implement the qp evacuation and the probing of $\nqp$ experimentally, we fabricate and measure a series of aluminum-based samples, cooled down in a dilution refrigerator reaching electronic base temperature $\tb\approx 50\,\mathrm{mK}$.

Here we show detailed results for two devices for confirmation of results, both with copper as the turnstile normal-metal island, separated by aluminum-oxide barriers from two aluminum leads. The samples were patterned by electron-beam lithography and metal deposition was done by multi-angle shadow electron-beam evaporation, see Supporting Information for further details on the fabrication process. One of the leads (left in Fig.~\ref{f1}(a)) is made wide to trap qps passively while the other (right) is long and narrow, this way promoting an excess qp population: due to the geometry, qp diffusion away from the turnstile junction and their subsequent relaxation is intentionally poor in this narrow lead~\cite{Knowles2012}. The samples were fabricated using standard electron-beam lithography and shadow mask techniques. The narrow lead is the $\mr{S}_1$ part of a $\sis$ Superconducting QUantum Interference Device (SQUID) whereas the fork-shaped electrode forms $\mr{S}_2$ with a higher superconducting gap. The different superconducting gaps are achieved by depositing a thicker aluminum layer as the turnstile lead ($d_1\approx 70\,\mr{nm}$) and a thinner one for the rest of the SQUID, ($d_2\approx 8\,\mr{nm}$)~\cite{Court2007}. A magnetic field perpendicular to the plane of the sample is applied to produce a flux $\Phi=\Phi_0/2$ (here $\Phi_0=h/(2e)$ is the magnetic flux quantum) in the SQUID loop and therefore suppress the supercurrent in the $\sis$ junctions so that subgap $\vsis$ can be applied~\cite{Ferguson2008} (see also the Supporting Information).

We characterize the samples by sweeping the gate voltage $\vg$ between the ``closed'' ($\nng=0$) and ``open'' ($\nng=0.5$) states (with $\nng=\cg\vg/e$, where $\cg$ is the gate capacitance), and stepping the SET bias voltage $\vb$. The $\sis$ junctions are first kept unbiased, named as the ``cooler-off'' case. The main parameters of the samples have been extracted from these measurements by fitting the measured current $I$ to current-voltage curves calculated with a master equation approach taking into account sequential tunnelling and Andreev reflections \cite{Hekking1994} (see for example Supporting Information and Ref. \onlinecite{Aref2011}). One of these fits can be seen in Fig.~\ref{f1}(c). For sample A (B) the total resistance of the tunnel barrier is $\rt=159.9\,\mathrm{k}\Omega$ ($63.0\,\mathrm{k}\Omega$), the charging energy of the island $\ec=0.95\Delta_1$ ($0.50\Delta_1$), the thickness of the N island $d=30\,\mathrm{nm}$ ($40\,\mathrm{nm}$) and the Dynes parameter \cite{Saira2010} $\eta=1.0\times 10^{-4}$ ($7.5\times 10^{-5}$). For both samples, the superconducting energy gap $\Delta_1$ of the leads is $180\,\mathrm{\mu eV}$, and the island's lateral dimensions $l\times w$ are $1\,\mathrm{\mu m}\times 100\,\mathrm{nm}$. See Supporting Information for further details on the employed model where it is explicit that the island temperature is calculated according to power balance and that the area of a single conduction channel $\left(\ach\right)$ can be extracted from these data. Yet it is more precise to extract $\ach$ from pumping measurements since the effects of the second-order tunnelling are more pronounced in that operation regime. Since qp diffusion is altered by the different lead shapes~\cite{Knowles2012} the DC fits were made by leaving temperature of the long lead as a free parameter and assuming the base temperature of $50\,\mr{mK}$ for the wide lead, estimating an electronic temperature $\sim 150\,\mr{mK}$ for the long lead. The values of $\ach$, $9.5\,\mathrm{nm}^2$ ($10\,\mathrm{nm}^2$) for sample A (B), are lower by about a factor of two compared to previously reported values \cite{Maisi2011}; yet the fits are relatively insensitive to the exact value of this parameter. We estimate these values from data with the cooler bias $\vsis$ near to certain optimal point, where the Andreev effects are more noticeable, as will be seen later. By analysing the differential conductance of the SQUID it was possible to estimate $\Delta_2$ to be $\sim 235\,\mr{\mu eV}$ and $\sim 240\,\mr{\mu eV}\pm 5\mr{\mu eV}$ for sample A and B, respectively.

\begin{figure}[ht!]
	\includegraphics[scale=0.422]{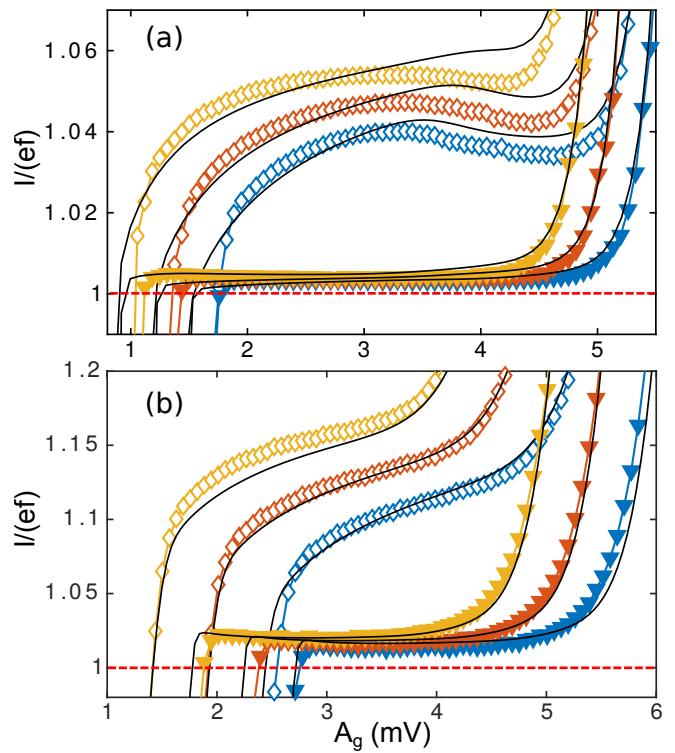}
\caption{Improvement of the $I=ef$ pumping plateaus by the $\sis$ cooler. Data for the cooler-off (open diamonds) and cooler-on (filled triangles) cases compared with a model based on a master equation approach (solid black lines). Blue, red and yellow curves correspond to $V_\mathrm{b}=120,\, 160,\, 200\, \mu\mathrm{V}$, respectively.  (a) Sample A operating at 80 MHz; in the cooler-on case the SQUID is biased at $60\,\mu\mathrm{V}$. (b) Sample B operating at 100 MHz; here in the cooler-on cases the SQUID is biased to $50\,\mu\mathrm{V}$.}
	\label{f2}
\end{figure}

The strong cooling by the superconducting junctions is evident in the turnstile operation of the SET shown in Fig.~\ref{f2}. In this regime a sinusoidal signal with an offset equivalent to $\nng=0.5$ is applied to the gate electrode and the amplitude of this signal is swept while the SET current is measured. The presented data shows these measurements zoomed to the first current plateau (see Fig.~\ref{f1}(d) for a typical measurement in a wider range) for both samples. The plots display the behaviour with the SQUID at zero bias and at a finite bias close to where optimum cooling is achieved. Besides displacing the plateau level from the expected value of $I=ef$, an elevated qp density in the leads close to the junctions makes the level depend on the SET bias voltage and hence the values of the current at the plateaus will spread \cite{Knowles2012}. In the two cases of Fig.~\ref{f2} the ``cooler-on'' $\left(\vsis\neq 0\right)$ curves show a smaller deviation from the expected current than the cooler-off ones. Additionally, the bias dependence of the plateau level is weaker with the cooler active. All this means that the density of excitations has been suppressed in the narrow lead by biasing the $\sis$ junctions. For these particular devices, the remaining deviation is explained by the relatively low $\ec<\Delta$ which promotes Andreev tunnelling, as well as by leakage current \cite{Aref2011} and remnant qp population as will be seen later.

\begin{figure}[ht!]
	\includegraphics[scale=0.425]{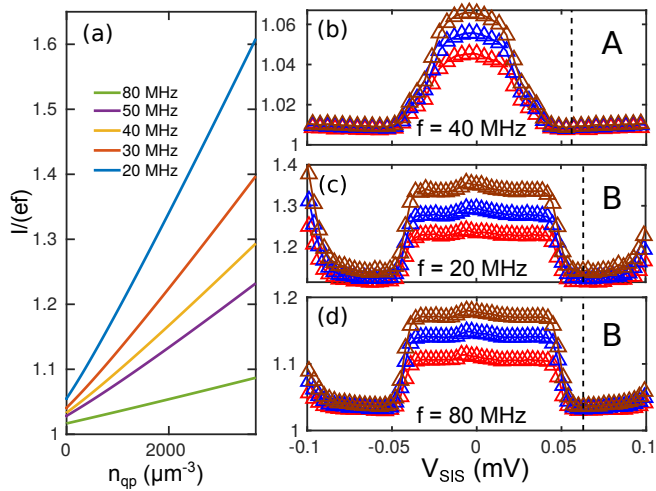} 
	\caption{Sensitivity of current plateau to qps. (a) Calculated curves showing how the current at the plateau varies with the qp density. The parameters of sample B were used in these calculations along with a bias $\vb=100\,\mr{\mu V}$ and a gate amplitude $A_\mr{g}=3.5\,\mr{mV}$. (b) Measured current at the plateau as a function of the bias applied to the $\sis$ cooler for device A at $f=40\,\mr{MHz}$. From bottom to top the curves correspond to $V_\mathrm{b}=120,\, 160,\,\mr{and}\, 200\, \mu\mathrm{V}$, respectively. (c) As in (b) for B at 20 MHz. (d) As in (c) for 80 MHz. Dashed black lines indicate the optimal biasing point of the cooler.}
	\label{f3}
\end{figure}

To quantify the reduction of the qp density, the pumping data at the plateau shown in Fig.~\ref{f2} are compared with a model (black solid curves) based on a master equation approach similar to that in the DC case. In the model the temperature of the island is varied at each amplitude value according to the proper power balance. Furthermore, since the used signal frequencies are high ($f>\tau_\mathrm{eff}^{-1}$, $\tau_\mathrm{eff}$ being the effective qp relaxation time), it is suitable to model the system as if the temperature of the superconductors was constant throughout the operation cycle~\cite{Maisi2013}. Additionally, electron-electron relaxation is assumed to be fast enough to avoid branch imbalance. With these assumptions the comparisons with the data are done with the temperature of the narrow lead as the only free parameter (device parameters are fixed by the DC measurements). As noted before, $\nqp$ is related to the effective electron temperature $T$ in the superconducting lead obtained by means of the thermal balance model. These simulations allow to deduce that the biasing of the $\sis$ junctions effectively cools the narrow long lead. Furthermore, the use of the pumping operation of a hybrid SET as a sensitive detector of qps is justified by these calculations. Fig.~\ref{f3}(a) shows a specific example of sample B biased at $\vb=100\,\mr{\mu V}$ and gate modulated with an amplitude of around $A_\mr{g}=3.5\,\mr{mV}$. There it is evident that the plateau level grows with the excitation density in the lead. The behaviour is nearly linear at high population but flattens for low values.

The behaviour of the qp density when the cooler bias changes is analysed by performing measurements where $\vsis$ is swept while the gate drive amplitude $A_\mr{g}$ is fixed to approximately middle of the first current plateau. Examples of such measurements are shown in Fig.~\ref{f3}(b) for device A, and in panels (c) and (d) for the B device. The plateau level and the spreading of them for different bias values of the SET, decrease until we reach $e\vsis\approx\pm|\Delta_2-\Delta_1|$~\cite{Melton1980} (indicated by a dashed line in panels b-d). At this point the temperature of the lead is at a minimum and it changes only weakly until a threshold at which it starts to grow rapidly. Therefore we can assert that when the cooler bias voltage reaches the optimum point the excitation population in the lead is at a minimum and at higher biases it starts to grow up first at a low rate and then rapidly when there are peaks in sub-gap current of the $\sis$ SQUID (see Supporting Information for further details). Similarly, biasing at $\vsis<0$ affects the system in the same way but this time by extracting hole-like excitations (consider the inset in Fig.~\ref{f1}(b) with the lower singularities aligned). We can also understand this by looking at the modelled cooling power of the $\sis$ junction \cite{Golubev2013} (see Supporting Information for details) when it begins to grow with $\vsis$ towards a sharp peak, where the lead temperature is minimum, and then it decreases dramatically, corresponding to the zone with finite but less efficient cooling. Finally, at $e\vsis\approx\Delta_2+\Delta_1$ qps are injected into the lead with the smaller gap. No sharp dip is observed in the measured $I$ vs. $\vsis$ curves at $e\vsis=|\Delta_2-\Delta_1|$, similar to earlier works on $\sis$ cooling~\cite{Manninen1999,Ferguson2008}. The singularity matching peak in the cooling power is likely washed out due to low-frequency noise in the $\sis$ cooler bias voltage, finite sub-gap density of states, and local inhomogeneities in the superconducting gap.

The influence of the $\sis$ cooler is qualitatively the same for both samples and independent of the operation frequency as well as of the SET bias, although the plateau levels are different, see the supplementary material for additional measurements to corroborate this fact. Note that this level approaches the ideal value but never reaches it even in the calculations for samples with low $\ec<\Delta$ (see Fig.~\ref{f3}(a)). In these devices with low $\ec$ we observe a clear crossover from excess qp-dominated to Andreev reflection-limited current quantization as the cooler is turned on.

\begin{figure}[ht!]
	\includegraphics[scale=0.48]{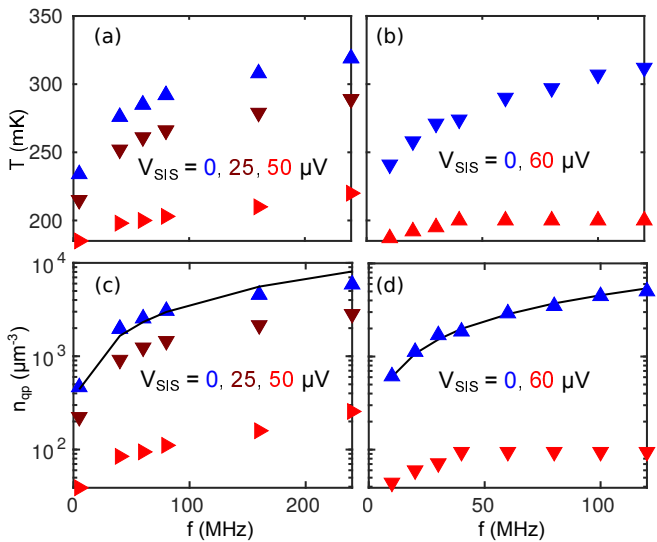}
	\caption{Qp density variation as a function of drive frequency. (a) Estimated narrow lead temperature during the turnstile operation as a function of the gate signal frequency for sample A and (b) for sample B. (c) Estimated qp density at the junction as a function of the pumping frequency for sample A and (d) B, corresponding to the data in panel (a) and (b), respectively. The black solid lines are fittings to Eq.~\eqref{e1}.}
	\label{f4}
\end{figure}

We estimate the lead temperature for a wide range of operation frequencies. Figs.~\ref{f4}(a) and (b) show that the lead temperature $T$ extracted from the fitting procedure, and in turn $\nqp$ (Figs.~\ref{f4}(c) and (d)) grow with increasing frequency also in the cooler-on case. However, in sample B the curve for the cooler-on case flattens at $f>40\,\mathrm{MHz}$ since the SET behaviour loses sensitivity to qp density in this operation regime (see Fig.~\ref{f3}(a)). For the near optimal cooler bias ($\vsis\approx 50\,\mu\mathrm{V}$ and $60\,\mu\mathrm{V}$ for A and B, respectively) the dependence of $\nqp$ versus the operation frequency is resemblant. This comes from the fact that qp transport is dominated by diffusion. There is dramatic drop in the qp density due to the biasing of the $\sis$ junctions. By linearly extrapolating the densities to the $f=0$ limit it is possible to see that there is a reduction between the ``off'' and ``on'' case by an order of magnitude also in this situation. We conclude that the cooler suppresses also the ever present excess qp population $n_\mr{qp,0}$, generated by non-equilibrium environment~\cite{Lambert2014,Lambert2017,Rossignol2019}, which in turn should diminish the subgap current in the DC regime. On the other hand it is seen that the biased $\sis$ junctions cannot totally remove this excess qp population in the leads. We expect that even more efficient evacuation can be achieved by placing the $\sis$ junction closer to the injection junction -- or even under it -- since the highest concentration of qps is near this point \cite{Rajauria2009,Riwar2016}, and further diffusion, recombination and phonon pair breaking, among other phenomena can thus be avoided.

We measured further reference samples with identical aluminum lead geometry but without $\sis$ SQUIDs and obtained the qp densities in them (for results see Supporting Information). The qp densities are similar to those shown in Fig.~\ref{f4} for the cooler-off case, and thus there is no significant influence of the unbiased cooler junctions or the presence of their biasing circuit on the relaxation of the non-equilibrium excitations. Hence the presented cooler-off densities should correspond approximately to densities in the case if the $\sis$ junctions were not present. Thus the diffusion model in \cite{Knowles2012}, which is based on heat conduction, without considering normal metal traps, should hold. Within this model the qp density is given by
\begin{equation}
\nqp=e^2\dfermi\pinj\dfrac{\rho_\mathrm{n}\ell}{\varpi\vartheta}\sqrt{\dfrac{\pi}{2\kb T\Delta_1}}+n_\mr{qp,0},
\label{e1}
\end{equation}
where $\ell,\,\varpi,\,\vartheta$ are the dimensions of the lead ($25\,\mathrm{\mu m}\times 100\,\mathrm{nm}\times 70\,\mathrm{nm}$), $\pinj$ is the injected power which can be approximated by $\Delta_1f$ in the driving regime of the experiment \cite{Knowles2012}. In Supporting Information we show that this is a good approximation. Notably, even at $f=100\,\mr{MHz}$, the injected power for an aluminium-based device with $\Delta_1 = 200 \mr{\mu eV}$ is only around $3.2\,\mr{fW}$, extractable by a sub-micron $\sis$ junction. In addition, $\rho_\mathrm{n}$ is the aluminium normal state resistivity. Based on measurements of a number of separate test structures, we estimate $\rho_\mr{n}\approx31\mr{\Omega nm}$ at 77 K (see the supplement for details). A few of these 4-probe structures were cooled down to 4.2 K where we estimate a further $10\%$ decrease of $\rho_\mr{n}$ below its 77 K value. The measured qp numbers are larger than those predicted by this model with no free parameters. Fitting Eq.~\eqref{e1} to the data using $\rho_\mathrm{n}$ as a free parameter yields $\rho_\mr{n}\sim 90\,\Omega\mathrm{nm}$. These fits are shown as black lines in panels (c) and (d) of Fig.~\ref{f4}. Further experiments are needed to understand the discrepancy, observed also in \cite{Mykkaenen2018} for devices with higher charging energy.

At $f=0$, $\vsis=0$, we estimate $n_\mr{qp,0}\approx 250\,\mr{\mu m^{-3}}$ for both samples, obtained from fits to dc IV characteristics of the turnstile. These background qp densities for the cooler-off case are roughly two orders of magnitude higher than those observed for highly-shielded SINIS turnstiles~\cite{Knowles2012}, or superconducting qubits and resonators~\cite{Visser2012,Visser2012a,Riste2013}. The excess in $n_\mr{qp,0}$ compared to Ref.~\onlinecite{Knowles2012} originates from the lack of a microwave-tight indium seal in the sample holders employed in these measurements and from the lack of any normal metal traps and restricted geometry -- narrow, thin, long -- of the S electrode with the $\sis$ junctions. To bring down $n_\mr{qp,0}$, a sample holder with a higher level of shielding can be utilized, and the $\mr{S_2}$ electrode can be fabricated with separate lithography and deposition steps that do not limit the junction geometry as with the multi-angle shadow evaporation technique where both NIS and $\sis$ junctions are created with a single mask.

The applicability of the qp extraction and detection techniques considered in this work extends beyond SINIS turnstiles and improving the accuracy of their pumped current. First, one can envision straightforward integration of $\sis$ coolers for instance with absorbers of kinetic inductance detectors, or into superconducting resonators analogously to NIS coolers~\cite{Tan2017}. When the bottom electrodes of the junctions are fabricated in a separate lithography and deposition step, their area and geometry can be adjusted at will. Furthermore, instead of tuning the gap by the film thickness, different superconducting materials can be utilized. Secondly, the SINIS turnstile, demonstrated here to function as a sensitive and direct qp probe, can be combined with various superconducting devices to measure the background qp density. This high-impedance, non-invasive probe can test the level of microwave shielding also in setups with sensitive non-superconducting devices. Finally, we have shown the driven hybrid turnstile to act as a highly controlled qp injector that could be used to investigate the qp sensitivity and qp trap efficiency of superconducting qubits and other devices.

In summary, we have been able to demonstrate active extraction of non-equilibrium excitations from a superconductor. A more than an order of magnitude reduction from values as high as $5.8\times 10^3\,\mu\mathrm{m}^{-3}$ to $260\,\mu\mathrm{m}^{-3}$ at the highest studied injection rate of $2.4\times 10^8\,\mr{s^{-1}}$ was achieved. Furthermore, in the limit of no qp injection ($f\rightarrow 0$) we find a similar reduction although a finite population of environmentally created excitations remains. Our work shows that the qp density in the superconducting electrodes can be controlled by active qp traps, here demonstrated for the first time with \textit{in-situ} control of qp injection.

\textbf{Supporting Information.} Calculations of the cooling power of $\sis$ junctions are contained here, as well as a model for the power transfer in a driven system. Additionally, details on fabrication and measurement procedures are given. Further details on the theoretical background and numerical calculations of SINIS SETs as well as results for additional and reference measurements are provided. Finally, details on the aluminum resistivity are given.

\section{AUTHOR CONTRIBUTIONS}
M.M.-S. performed simulations, analysed the data and performed the resistivity measurements. J.T.P. fabricated the devices and performed the rest of the experiments. The calculation of the injected power was done by J.P.P and M.M.-S. The manuscript was written by M.M.-S. with important input from J.T.P. and J.P.P. M.M.-S. and J.T.P. contributed equally to the research.

\section{COMPETING INTERESTS}
The authors declare no competing interests

\section{ACKNOWLEDGEMENTS}
We acknowledge E. T. Mannila, D. Golubev, B. Karimi, V. F. Maisi and I. M. Khaymovich for useful discussions. This research made use of Otaniemi Research Infrastructure
for Micro and Nanotechnologies (OtaNano)  and its Low Temperature Laboratory, which is part of the European Microkelvin Platform (EMP). This work is funded through Academy of Finland Grants No. 297240, No. 312057, and No. 303677.

\end{document}


\title{Supporting Information\\ Active quasiparticle suppression in a non-equilibrium superconductor}
\author{Marco Marín-Suárez}
\author{Joonas T. Peltonen}
\author{Jukka P. Pekola}
\affiliation{Pico group, QTF Centre of Excellence, Department of Applied Physics, Aalto University, FI-000 76 Aalto, Finland}
\begin{abstract}
\end{abstract}

\maketitle
\section{COOLING POWER OF A JOSEPHSON JUNCTION}

\begin{figure}[ht!]
\includegraphics[scale=0.5]{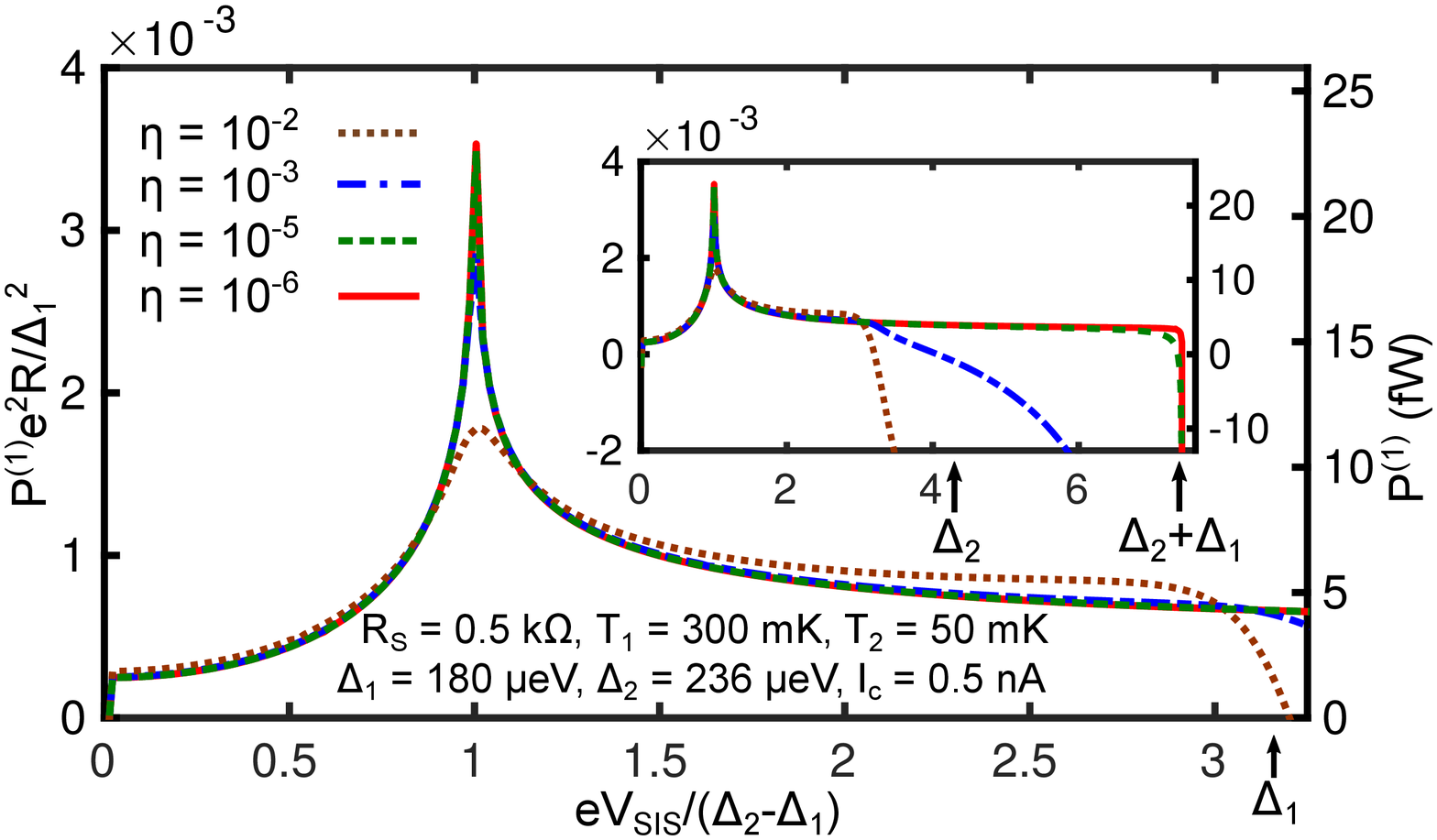}
\caption{Normalized cooling power of a $\sis$ tunnel junction for different Dynes parameters. The junction parameters used in the calculation are given in the figure. In addition, to demonstrate practical magnitude of $P^{(1)}$ we set $R=5\,\mr{k\Omega}$, which is typical for our devices, to obtain the scale on the right hand side. The inset shows calculations in a wider voltage range showing the onset of strong heating. The bias points corresponding to $\Delta_1$, $\Delta_2$ and $\Delta_1+\Delta_2$ are indicated by arrows.}
\label{f8}
\end{figure}

A simplified model of a resistively shunted superconducting junction can be developed for calculating the cooling power of one of the superconductors~\cite{Golubev2013}. The system consists of a Josephson junction composed of a tunnel barrier between  two leads with superconducting gaps $\Delta_i$ and temperatures $T_i$ with $i=1,\, 2$. The junction is connected in series to a resistance $R_\mathrm{S}$ and fed by a bias $V_\mathrm{SIS}$. The system dynamics is given by
\begin{equation}
\begin{split}
&V_\mathrm{SIS}=I_\mr{SIS}R_\mathrm{S}+\dfrac{\hbar\dot{\varphi}}{2e},\\
&I_\mr{SIS}=I_\mathrm{c}\sin{\varphi},
\end{split}
\label{e20}
\end{equation}
where $e$ is the elemental charge and $I_\mathrm{c}$ is the critical current of the junction. For a single junction, $\Ic$ can be approximated by~\cite{Ambegaokar1963}
\begin{equation}
\Ic=\dfrac{1}{eR}\int_{\Delta_1}^{\Delta_2}dE\dfrac{\Delta_1\Delta_2\left[1-2f_1\left(E\right)\right]}{\sqrt{E^2-\Delta_1^2}\sqrt{E^2-\Delta_2^2}}.
\end{equation}
Here, $f_1$ denotes the distribution function of the superconductor 1 (often taken to be the Fermi-Dirac function), whereas $R$ is the normal-state tunnel resistance.

In the experiments, instead of an individual $\sis$ junction we use a DC superconducting quantum interference device (SQUID) with two junctions in parallel to be able to minimize $\Ic$ by a magnetic flux $\Phi$. Neglecting the inductance of the loop, the critical current is given by $\Ic=\sqrt{I_\mr{c,L}^2+I_\mr{c,R}^2+2I_\mr{c,L}I_\mr{c,R}\cos\left(2\pi\Phi/\Phi_0\right)}$, where $I_\mr{c,L/R}$ are the critical currents of the individual junctions and $\Phi_0=h/(2e)$ is the flux quantum. We apply an external magnetic flux $\Phi=\Phi_0/2$ through the SQUID loop so that $\Ic=|I_\mr{c,L}-I_\mr{c,R}|$ is at a minimum, due to typical junction size asymmetry at a value around $0.5\,\mr{nA}$.

Suppressing $\Ic$ allows us to voltage bias the SQUID at subgap voltages, essential for the $\sis$ cooling. In the simplest approximation, we neglect the cooling power of the SQUID junction that is further away from the SINIS turnstile, and model the SQUID as a single junction with the suppressed $\Ic$.
\begin{figure}[ht!]
\includegraphics[scale=0.75]{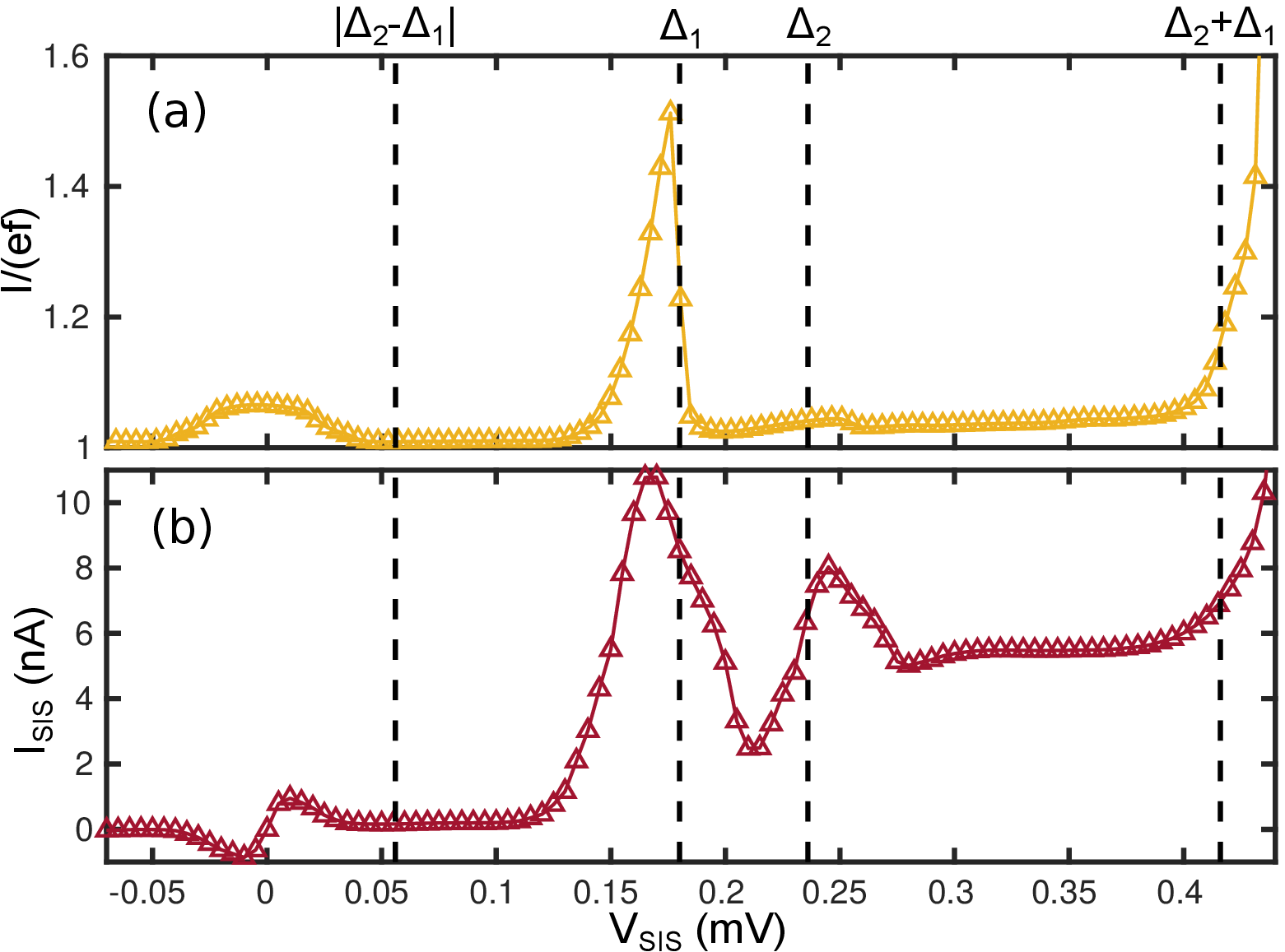}
\caption{(a) Pumped current of sample A at a constant drive amplitude on the $I=ef$ plateau. It is measured at $f=40\,\mr{MHz},\, \vb = 200\,\mr{\mu V}$ and plotted as a function of the bias applied to the $\sis$ cooler over a wide range. (b) Sample A SQUID current-voltage curve at $\Phi\approx \Phi_0/2$. Dashed vertical lines indicate the bias voltages $eV_\mr{SIS}=|\Delta_2-\Delta_1|,\,\Delta_1,\,\Delta_2,$ and $\Delta_2+\Delta_1$.}
\label{f11}
\end{figure}

Combining Eqs. \eqref{e20} one obtains
\begin{equation}
\dfrac{1}{R_\mathrm{S}}\dfrac{\hbar \dot{\varphi}}{2e}+\Ic\sin{\varphi}=\dfrac{V_\mathrm{SIS}}{R_\mathrm{S}}.
\end{equation}
In a Josephson junction $\dot{\varphi}=0$ when $V_\mathrm{SIS}<\Ic R_\mathrm{S}$ and $\varphi=\arcsin{\left(V_\mathrm{SIS}/\Ic R_\mathrm{S}\right)}$. Otherwise
\begin{equation}
\begin{split}
\cos{\varphi(t)}&=\dfrac{\sqrt{V_\mathrm{SIS}^2-\Ic^2R_\mathrm{S}^2}\sin\left(\frac{2e}{\hbar}t\sqrt{V_\mathrm{SIS}^2-\Ic^2R_\mathrm{S}^2}\right)}{V_\mathrm{SIS}-\Ic R_\mathrm{S}\cos\left(\frac{2e}{\hbar}t\sqrt{V_\mathrm{SIS}^2-\Ic^2R_\mathrm{S^2}}\right)}, \\
\dfrac{\hbar\dot{\varphi}(t)}{2e}&=\dfrac{V_\mathrm{SIS}^2-\Ic^2R_\mathrm{S}^2}{V_\mathrm{SIS}-\Ic R_\mathrm{S}\cos\left(\frac{2e}{\hbar}t\sqrt{V_\mathrm{SIS}^2-\Ic^2R_\mathrm{S}^2}\right)}.
\end{split}
\end{equation}

The power transferred from the lead $i=1$ is written as $P^{(1)}(t)=P^{(1)}_\mathrm{qp}(t)+P^{(1)}_{\mr{J}}(t)$, where $P^{(1)}_\mathrm{qp}(t)$ is the power transmitted by quasiparticles (qp's) given by
\begin{equation}
P^{(1)}_\mathrm{qp}\left(V(t)\right)=\dfrac{1}{e^2R}\int{dEn_1\left(E-eV(t)\right)n_2\left(E\right)\left(E-eV(t)\right)\left[f_1\left(E-eV(t)\right)-f_2\left(E\right)\right]}.
\label{e30}
\end{equation}
In \eqref{e30}, $n_i$ denotes the density of qp states in lead $i$, modeled by the broadened BCS DoS
\begin{equation}
n_i\left(E\right)=\left|\mathfrak{Re}\left(\dfrac{E/\Delta_i+i\eta}{\sqrt{\left(E/\Delta_i+i\eta\right)^2-1}}\right)\right|.
\label{e11}
\end{equation}
Here, $\eta$ is the Dynes parameters which models subgap leaks~\cite{Dynes1978,Pekola2010}. Additionally, $f_i$ denote the distribution functions. On the other hand the Josephson contribution \cite{Golubev2013} $P^{(1)}_\mr{J}(t)$ can be written as $P^{(1)}_{\cos}(t)\cos{\varphi}+P^{(1)}_{\sin}(t)\sin{\varphi}$, where
\begin{equation}
P^{(1)}_{\cos}\left(V(t)\right)=-\dfrac{1}{e^2R}\int{dEn_1\left(E-eV(t)\right)n_2\left(E\right)\dfrac{\Delta_1\Delta_2}{E}\left[f_1\left(E-eV(t)\right)-f_2\left(E\right)\right]}.
\end{equation}
When averaging the power in time the sine contribution vanishes and the power extracted from the lead $i=1$ is
\begin{equation}
P^{(1)}=\left\langle P^{(1)}_\mr{qp}\left(\dfrac{\hbar\dot{\varphi}(t)}{2e}\right)+P^{(1)}_{\cos}\left(\dfrac{\hbar\dot{\varphi}(t)}{2e}\right)\cos{\varphi(t)}\right\rangle_t.
\label{e5}
\end{equation}
When the junction is biased above critical current $\left(\vsis >\Ic R_\mr{S}\right)$ the cosine component vanishes in average. If $\Delta_2 >\Delta_1$ the power is positive and the average is peaked at $eV_\mathrm{SIS}=\pm |\Delta_1-\Delta_2|$, as shown in Fig.~\ref{f8} for various Dynes parameters. Notice how the peak gets smeared at large $\eta$. Additionally, observe that for large Dynes parameters the power becomes negative at biases closer to $\Delta_1$, indicating injection of qp's to $\mr{S_1}$. For low leakage cases this heating does not start until close to $e\vsis >\Delta_1 +\Delta_2$.

Figure~\ref{f11}(a) shows the large-scale behavior of the current at the plateau against $V_\mr{SIS}$ for sample A at $f=40\,\mr{MHz}$ and $\vb=200\,\mr{\mu V}$. As already shown in the main manuscript the lowest current is found around $e\vsis=|\Delta_2-\Delta_1|$ in agreement with the model. Here, the response is smeared likely by subgap leakage, local inhomogeneities of the gap, and noise in the bias voltage. Further discrepancies with the model such as the strong heating around $\Delta_1$ correspond to peaks in subgap current of the SQUID as can be seen in Fig.~\ref{f11}(b) and are not caught by this simple model. Furthermore, a much weaker heating peak at $e\vsis\approx \Delta_2$ is present. Finally, observe that when the current-voltage curve starts to become linear at $e\vsis >\Delta_1+\Delta_2$ there is a strong heating since qp's are now injected from $\mr{S_2}$ to $\mr{S_1}$ in accordance with the basic model. To simulate the subgap peaks, expressions for the cooling power should be developed and solved in a framework that includes multiple Andreev reflections, analogous to calculations of the electric current e.g. in Ref.~\onlinecite{Hurd1996}. In a more detailed model also the temperature gradient along the $\mr{S_1}$ electrode and the electromagnetic environment of the $\sis$ junction should be considered.

\section{POWER TRANSFER IN A DRIVEN SYSTEM}

The process of transferring electrons from the island to the leads can be regarded as a jump stochastic process whose probability of jumping is governed by the equation
\begin{equation}
\dfrac{d}{dt}p(t)=-\Gamma(t)p(t).
\label{e3}
\end{equation}
The reverse process is assumed to vanish and $\Gamma$ is the rate, which in the slow driving regime depends on the driving as $\Gamma=\Gamma_0e^{-\tilde{\beta}(1-\varepsilon)}$. For the present case, $\Gamma_0=\frac{\Delta}{e^2R}$, with $R$ the tunnel resistance and $\tilde{\beta}=\frac{\Delta}{k_\mathrm{B}T}$ ($k_\mathrm{B}$ is the Boltzmann constant, $T$ is the temperature and $\Delta$ the superconducting gap). Furthermore, $\varepsilon (t)=2E_\mr{c}/\Delta\left(n_\mr{g}(t)-1/2\right)$ is the energy of the transferred electron normalized to the energy gap, where $E_\mr{c}$ is the charging energy of the island and $n_\mr{g}$ is the charge induced by the driving signal.

\begin{figure}[ht!]
\includegraphics[scale=0.7]{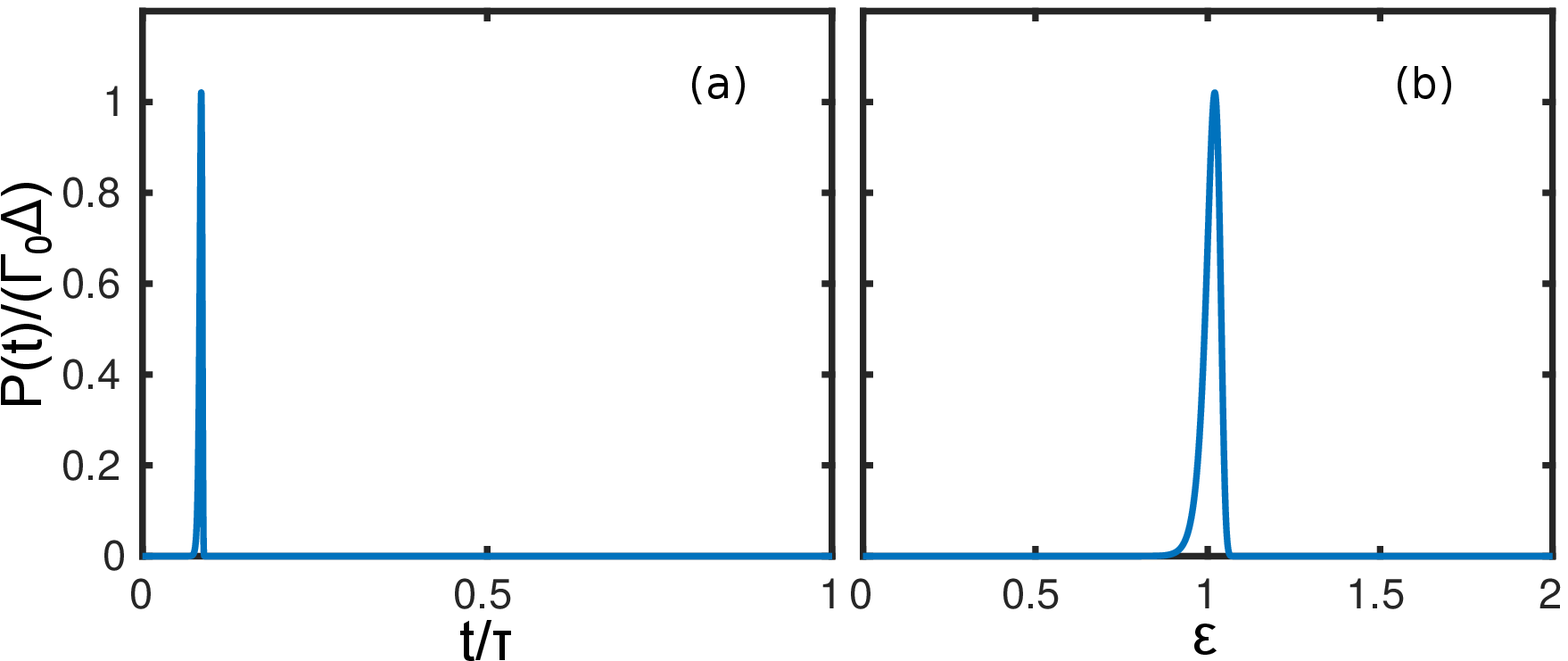}
\caption{Dynamics of a driven system with $n_\mathrm{g}(t)=0.5+\sin\left(2\pi ft\right)$, $f=50\,\mathrm{MHz}$, $E_\mathrm{c}=\Delta$, $\tilde{\beta}=50$ and $\Gamma_0=10\,\mathrm{GHz}$. (a) Power as function of time. (b) Power as a function of the driven energy, notice how it sharply peaks when $\varepsilon=1$.}
\label{f9}
\end{figure}

The solution of Eq. \eqref{e3} is
\begin{equation}
p(t)=\exp\left(-\Gamma_0\int_0^te^{-\tilde{\beta}\left(1-\varepsilon(x)\right)}dx\right).
\end{equation}
Now, notice that the power transfer is given by $P(t)=\dot{E}=\Delta\varepsilon\Gamma(t)p(t)$. The average transmitted power during one drive period of length $\tau=1/f$ is then given by
\begin{equation}
\langle P\rangle =\dfrac{\Delta}{\tau}\int_{0}^{\tau} \Gamma_0\varepsilon e^{-\tilde{\beta}\left(1-\varepsilon(t)\right)}p(t)dt.
\label{e4}
\end{equation}
In the conditions of interest $\tilde{\beta}\gg 1$ and $\Gamma\gg f$, with $f$ the frequency of the driving. The integral is thus $\sim 1$ for a general periodic driving and $\langle P\rangle\approx \Delta f$. A specific case of the dynamics of the instantaneous power transfer is shown in Fig.~\ref{f9}(a). Notice how it sharply peaks at a specific time making its integral nearly one and the energy transmitted during one cycle $\Delta$. Fig.~\ref{f9}(b) depicts how the power evolves with the driven electron energy for a quarter of cycle. Observe how almost all the power is transferred when $\varepsilon(t)=1$ and $\Delta=2E_\mathrm{c}\left(n_\mathrm{g}(t)-1/2\right)$.

\section{METHODS}
\subsection{Fabrication}

The samples were fabricated on 4-inch silicon substrates covered by 300 nm thermal silicon oxide. The process is based on electron beam lithography (EBL, Vistec EBPG5000+ operating at 100 kV) and multiangle shadow deposition of metals in an electron-beam evaporator. First, a set of bottom gates (visible e.g. in Fig.~1(a) of the main text) and large ground plane electrodes were formed by depositing a 2 nm titanium adhesion layer, 30 nm gold, and another 2 nm Ti protection layer through a mask defined in a single layer positive resist (Allresist AR-P 6200). After conventional liftoff, the gate and groundplane electrodes are covered by a 50 nm insulating layer of $\mr{Al_2O_3}$ grown by atomic layer deposition (ALD). Next, another round of EBL and metal evaporation (2 nm Ti followed by 30 nm AuPd) is used to form bonding pads and coarse electrodes for connecting to the tunnel junctions, to be patterned in the third and final lithography step. For defining the $\sis$ and NIS junctions, a Ge-based hard mask is used~\cite{Pekola2013}. It is composed of a 400 nm sacrificial layer of P(MMA-MAA) copolymer, covered by 22 nm Ge also deposited by e-gun evaporation, and finally a thin (approximately 50 nm) layer of PMMA on top. Before development of the Ge mask, the wafer was cleaved into smaller, typically 1 cm x 1 cm chips. The pattern defined into the PMMA resist is transferred to the Ge layer by $\mr{CF}_4$ reactive ion etching (RIE). Subsequently, an undercut profile is formed in the copolymer layer by oxygen plasma etching in the same RIE system.

The tunnel junction deposition starts by evaporating a 13 nm layer of Al at a substrate tilt angle of approximately $55^{\circ}$, resulting in an approximately 8 nm film that forms the fork-shaped $\mr{S_2}$ electrode of the SQUID. Right after deposition, this film is oxidized \emph{in situ} in the evaporator (static oxidation with typically 2 mbar of pure $\mr{O}_2$ in the chamber for two minutes). A subsequent deposition of 70 nm aluminum at normal incidence angle (zero tilt) completes the $\sis$ SQUID and forms the $\mr{S_1}$ electrodes of the turnstile. Finally, the SINIS turnstile and the whole device are completed by a second oxidation (typically 1 mbar $\mr{O}_{2}$ for one minute) and depositing 50 nm of copper under a tilt angle of $39.5^{\circ}$ (in the opposite direction compared to the $\mr{S_2}$ layer), resulting in an approximately 40 nm thick Cu island. For measurements at cryogenic temperatures, a chip with an array of 3-by-3 devices is cleaved to fit a custom-made chip carrier and electrically connected to it by Al wire bonds. Due to the limited number of measurement lines, typically 2--3 devices from the chip are selected to be characterized at low temperatures.

\subsection{Measurements}
Experiments were carried out in a custom-made plastic dilution refrigerator with base temperature of about 50 mK. Conventional cryogenic signal lines (resistive twisted pairs between room temperature and the 1 K pot flange, followed by at least 1 m Thermocoax cable as a microwave filter to the base temperature), connect the bonded chip to a room-temperature breakout box. The rf line for applying the drive signal to the turnstile gate consists of a stainless steel coaxial cable down to 4.2 K, a 20 dB attenuator in the liquid helium bath, followed by a feedthrough into the inner vacuum can of the cryostat. Inside the cryostat, the rf signal is carried by a continuous superconducting NbTi coaxial cable from the 1 K stage down to the sample carrier. Magnetic field for the $\sis$ SQUID was applied by current biasing a superconducting coil placed on the exterior of the inner vacuum can. Current and voltage biases were realized by programmable voltage sources and function generators. Current amplification was achieved by room-temperature low-noise transimpedance amplifiers (FEMTO Messtechnik GmbH, models DDPCA-300 and LCA-2-10T). The curves of the pumped single-electron current were typically iterated at least 10 times and averaged accordingly, neglecting those repetitions during which an offset charge jump had occurred. Amplifier offset currents were subtracted by comparing the pumping curves with their counterparts measured under source-drain bias of opposite polarity.

\section{NUMERICAL CALCULATIONS}

The current-voltage characteristics of a SET can be modelled from a stochastic master equation on the charging events $n$, where $n$ designates the charge state of the island. The probability $p\left(n\right)$ evolves as \cite{Kulik1975,Likharev1985,Averin1986,Saira2013}
\begin{equation}
\dfrac{d}{dt}p\left(n,t\right)=\sum_{n'\neq n}{\gamma_{n'n}p\left(n',t\right)-\gamma_ {nn'}p\left(n,t\right)},
\label{e1}
\end{equation}
where $\gamma_{nn'}$ is the total transition rate from the $n$ state of the island to the $n'$ state. Furthermore, in the steady state $dp\left(n,t\right)/dt=0$ and since the charging events $n$ are discrete one can express Eq. \eqref{e1} as the matrix equation
\begin{equation}
A\mathbf{p}=\mathbf{0},
\end{equation}
where $p_i=p(n_i)$, $A_{ii}=-\sum_{j\neq i}\gamma_{ij}$ and $A_{ij}=\gamma_{ij}$ for $i\neq j$. Generalizing for two leads, as is the case of an SET, the transition rates are given by $\gamma_{nn'}=\Gamma_{n\rightarrow n'}^l\left(\delta E\right)+\Gamma_{n\rightarrow n'}^r\left(\delta E\right)$. Here, $\Gamma$ is the transition rate between individual charging states $n$ and $n'$, $l$ designates events between the left lead and the island and $r$ similarly stands for the right lead. Finally, $\delta E$ is the energy change of the process given by
\begin{align}
\delta E^{\pm,l/r}_{1e}\left(n\right)&=\mp 2E_\mr{c}\left(n-n_\mr{g}\pm 0.5\right)\pm eV_{l/r} \label{e6}\\
\delta E^{\pm,l/r}_{2e}\left(n\right)&=\mp 4E_\mr{c}\left(n-n_\mr{g}\pm 1\right)\pm 2eV_{l/r}, \label{e7}
\end{align}
for single-electron $\left(1e\right)$ and two-electron $\left(2e\right)$ tunnelling, respectively. Here $V_l=\kappa_lV_\mr{b}$ and $V_r=-\kappa_rV_\mr{b}$, $\kappa_{l/r}$ is the ratio between the junction capacitance and the total capacitance, $V_\mr{b}$ is the bias voltage applied between the leads of the transistor. In Eqs.~\eqref{e6} and \eqref{e7} $n$ is the initial island excess charge, $n_\mr{g}$ is the charge number induced by the gate voltage, $E_\mr{c}$ is the charging energy and $+\,(-)$ designates tunnelling to (from) the island.

The explicit expression for the rates depends on the specific system and on the transition processes taken into account. Since here we consider transitions in NIS junctions and only single-electron and two-electron Andreev processes are taken into account, the rates are given by
\begin{equation}
\Gamma^{l/r}_{n\rightarrow n\pm 1}\left(\delta E\right)=\dfrac{1}{e^2\rt}\int dE{n_s\left(E\right)\left(1-f_\mr{N}\left(E+\delta E\right)\right)f_\mr{S}\left(E\right)} \label{e8}
\end{equation}
for $1e$ tunnelling, and
\begin{equation}
\begin{split}
&\Gamma^{l/r}_{n\rightarrow n\pm 2}\left(\delta E\right)=\dfrac{\hbar\Delta^2}{16\pi e^4\rt^2\mathcal{N}}\int dEf_\mr{N}\left(E-\delta E/2\right)f_\mr{N}\left(-E-\delta E/2\right)\times\\
&\left|a\left(E+E_c-i\delta/2\right)+a\left(-E+E_c-i\delta/2\right)\right|^2 \label{e9}
\end{split}
\end{equation}
for Andreev tunnelling, and $\delta E$ is the corresponding energy change from Eqs.~\eqref{e6} or \eqref{e7}, respectively.

In Eqs.~\eqref{e8} and \eqref{e9}, $\Delta$ is the superconducting gap of the leads, $\rt$ is the tunnel resistance of the junction involved in the event either left or right and $\mathcal{N}$ is the number of conduction channels which can be written as $A/A_{\mathrm{ch}}$ with $A$ being the junction area and $A_{\mathrm{ch}}$ is the area of an individual channel. The term $\delta$ takes into account the energy of the intermediate (single-electron tunnelling) state which has a finite lifetime \cite{Averin2008}. Additionally, $f_\mr{N}$ is the Fermi-Dirac distribution of the normal-metal island, $f_\mr{S}$ is that for the superconducting lead involved in the tunnelling event and $n_s$ is the superconducting density of states given by Eq.~\eqref{e11}. Furthermore,
\begin{equation}
a\left(x\right)=\dfrac{1}{\sqrt{x^2-\Delta^2}}\ln\left(\dfrac{\Delta-x+\sqrt{x^2-\Delta^2}}{\Delta-x-\sqrt{x^2-\Delta^2}}\right).
\end{equation}

Once the probability vector $\mathbf{p}$ is obtained, the current through the SET can be calculated as $I=\mathbf{b}\cdot\mathbf{p}$ with $b_i=e\left(\Gamma_ {i\rightarrow i+1}^l-\Gamma_{i\rightarrow i-1}^l\right)+2e\left(\Gamma_ {i\rightarrow i+2}^l-\Gamma_{i\rightarrow i-2}^l\right)$. In order to get realistic results the temperature change of the island has to be taken into account. To do this one calculates the power transferred to the normal island in a single electron event as
\begin{equation}
\dot{Q}^{\mathrm{N},l/r}_{n\rightarrow n\pm 1}\left(\delta E\right)=\dfrac{1}{e^2\rt}\int{dEEn_s\left(E-\delta E\right)f_\mr{N}\left(E\right)\left(1-f_\mr{S}\left(E-\delta E\right)\right)},
\end{equation}
where N refers to normal metal and $\delta E$ is again the related energy change cost from Eq. \eqref{e6}. The total power transferred to the island is calculated as $\dot{Q}=\mathbf{q}\cdot\mathbf{p}$ with $q_i=\dot{Q}^\mathrm{N}_{i\rightarrow i+1}+\dot{Q}^\mathrm{N}_{i\rightarrow i-1}$, where $\dot{Q}^\mathrm{N}_{i\rightarrow i\pm 1}=\dot{Q}^{\mathrm{N},r}_{i\rightarrow i\pm 1}+\dot{Q}^{\mathrm{N},l}_{i\rightarrow i\pm 1}$. The heat flow to the phonons in the N island is governed by $\dot{Q}_\mathrm{e-ph}=\mathcal{V}\Sigma\left(T_\mathrm{N}^5-T_\mathrm{b}^5\right)$ where $\mathcal{V}$ is the volume of the island, $\Sigma$ is the electron-phonon coupling constant ($\approx 2.5\times 10^9\,\mr{WK^{-5}m^{-3}}$ for copper, close to previously measured values~\cite{Giazotto2006,Saira2007}), $T_\mathrm{N}$ is the electron temperature of the island and $T_\mathrm{b}$ is the temperature of the phonon bath which is considered to be the same as the cryostat mixing chamber temperature. An additional power transfer due to Andreev reflection is considered in the form of Joule heat, i.e. $\dot{Q}_\mr{A}=\left\langle I_\mathrm{A}\right\rangle V_\mathrm{b}$ where $I_\mathrm{A}$ is the current due only to Andreev events and $V_\mathrm{b}$ is the applied bias voltage. Finally, the heat balance is $\dot{Q}_\mathrm{e-ph}=\dot{Q}+\dot{Q}_\mathrm{A}$. This condition is used to solve for $T_\mathrm{N}$.

A similar method can be applied when a periodic gate drive is applied to the system~\cite{Saira2013}. If the period of this driving is $\tau$ then it is valid to assume that the steady state probability satisfies $p\left(t\right)=p\left(t+\tau\right)$. In order to solve for the probability the cycle is discretized in $m$ intervals of length $\Delta t=\tau/m$, next the matrix $A\left(k\Delta t\right)=A_k$ is calculated for each interval as well as the $\mathbf{b}_k$ vector. If we now build the propagator
\begin{equation}
U\left(\tau\right)=\prod_{k=1}^m\exp{\left(\Delta tA_k\right)},
\end{equation}
then the initial probability is given by
\begin{equation}
U\left(\tau\right)\mathbf{p}(0)=\mathbf{p}(0),
\label{e2}
\end{equation}
since for a periodic driving $\mathbf{p}(0)=\mathbf{p}(\tau)$.

A more useful form of the propagator is given by
\begin{equation}
\tilde{U}\left(\tau\right)=\prod_{k=1}^m\exp{\left(\Delta t\tilde{A}_k\right)},
\end{equation}
where
\begin{equation}
\tilde{A}_k=
\begin{bmatrix}
A_k & \mathbf{0}\\ \mathbf{b}_k^\mr{T} & 0
\end{bmatrix}.
\label{e10}
\end{equation}

In Eq.~\eqref{e10} $\mathbf{0}$ is an appropriate vector of zeros. This way Eq.~\eqref{e1} can be reformulated in terms of the augmented rate matrix $\left(\tilde{A}_k\right)$ and a new probability vector $\mathbf{\tilde{p}}\left(t\right)=\left[\mathbf{p}\left(t\right)\quad\langle q\rangle\right]^\mr{T}$, where $\langle q\rangle$ is the average charge transferred during one cycle. The final propagator can be decomposed as
\begin{equation}
\tilde{U}\left(\tau\right)=
\begin{bmatrix}
U & \mathbf{0}\\ \mathbf{U}_\mr{b} & 0
\end{bmatrix}.
\end{equation}
Then, Eq.~\eqref{e2} is reformulated in terms of this new propagator and the average charge is obtained as $\langle q\rangle=\mathbf{U}_\mr{b}\cdot \mathbf{p}(0)$. The average current pumped is $I=\langle q\rangle/\tau$. To complement this calculation the temperature of the island is calculated self-consistently, the same power balance is applied and the average heat transferred is now calculated with a similar propagator. However, now $\mathbf{b}_k$ has to be replaced by $\mathbf{q}_k$ and therefore $\dot{Q}=\langle Q\rangle/\tau$.

\section{ADDITIONAL MEASUREMENTS}
\subsection{Sample with different $\Delta_1$ and $\Delta_2$}
\begin{figure}[ht!]
\includegraphics[scale=0.5]{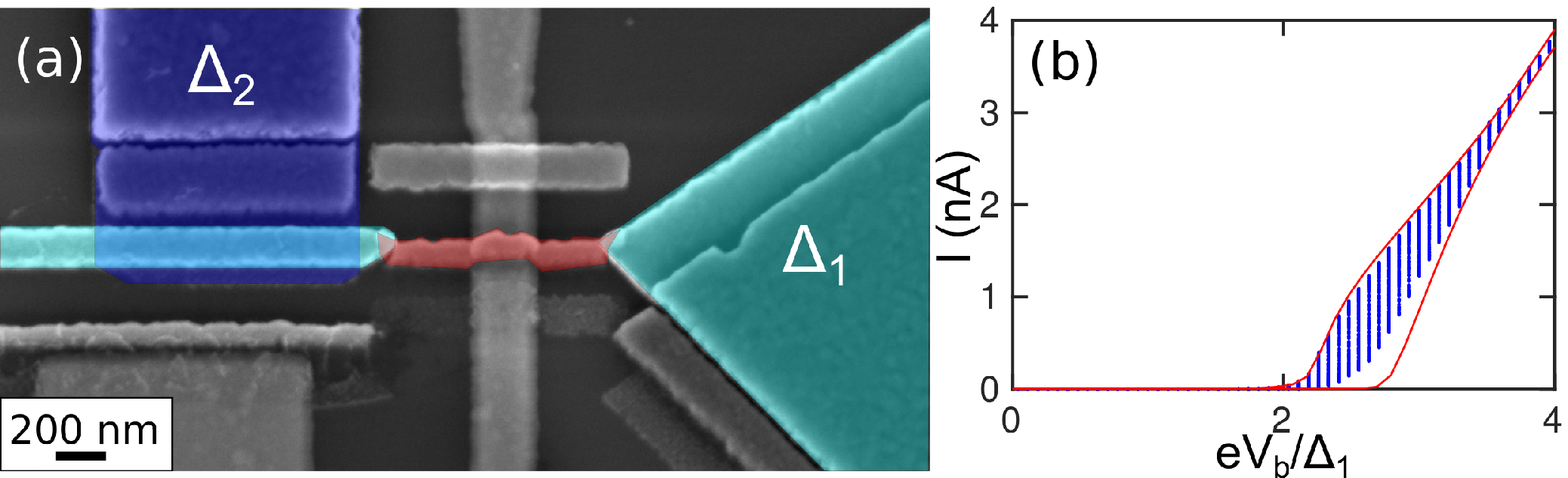}
\caption{Sample with different superconducting gaps (a) SEM micrograph of the measured sample with smaller difference $|\Delta_2-\Delta_1|$. Blue designates superconductor and red normal metal. Different tones of blue mean different gap superconductors. (b) IV curve in the DC regime. Blue dots are measured data and red lines are calculated curves from which device parameters are estimated.}
\label{f3}
\end{figure}

A third sample similar to A and B (see Fig.~\ref{f3}(a)) was measured. However, in this sample the turnstile leads are thinner $\left(d_1\approx 30\,\mathrm{nm}\right)$ and therefore their superconducting gap is higher than in the main samples A and B. Additionally, the superconductor $\mr{S_2}$ is thicker than in the main samples $\left(d_2\approx 17\,\mathrm{nm}\right)$. The parameters of the SET, determined by comparison with simulations (see Fig.~\ref{f3}(b)), are $\Delta_1=205\,\mathrm{\mu eV}$, $E_\mr{c}=0.53\Delta_1$, $\rt=97\,\mathrm{k\Omega}$, $A_\mathrm{ch}=5.5\,\mathrm{nm^2}$, and $\eta=2.75\times 10^{-5}$. The NIS junctions of this structure turned out to be asymmetric with a capacitance ratio of $C_r/C_l\approx 0.2$ where $l$ is for the left and $r$ for the right junction.

\begin{figure}[ht!]
\includegraphics[scale=0.5]{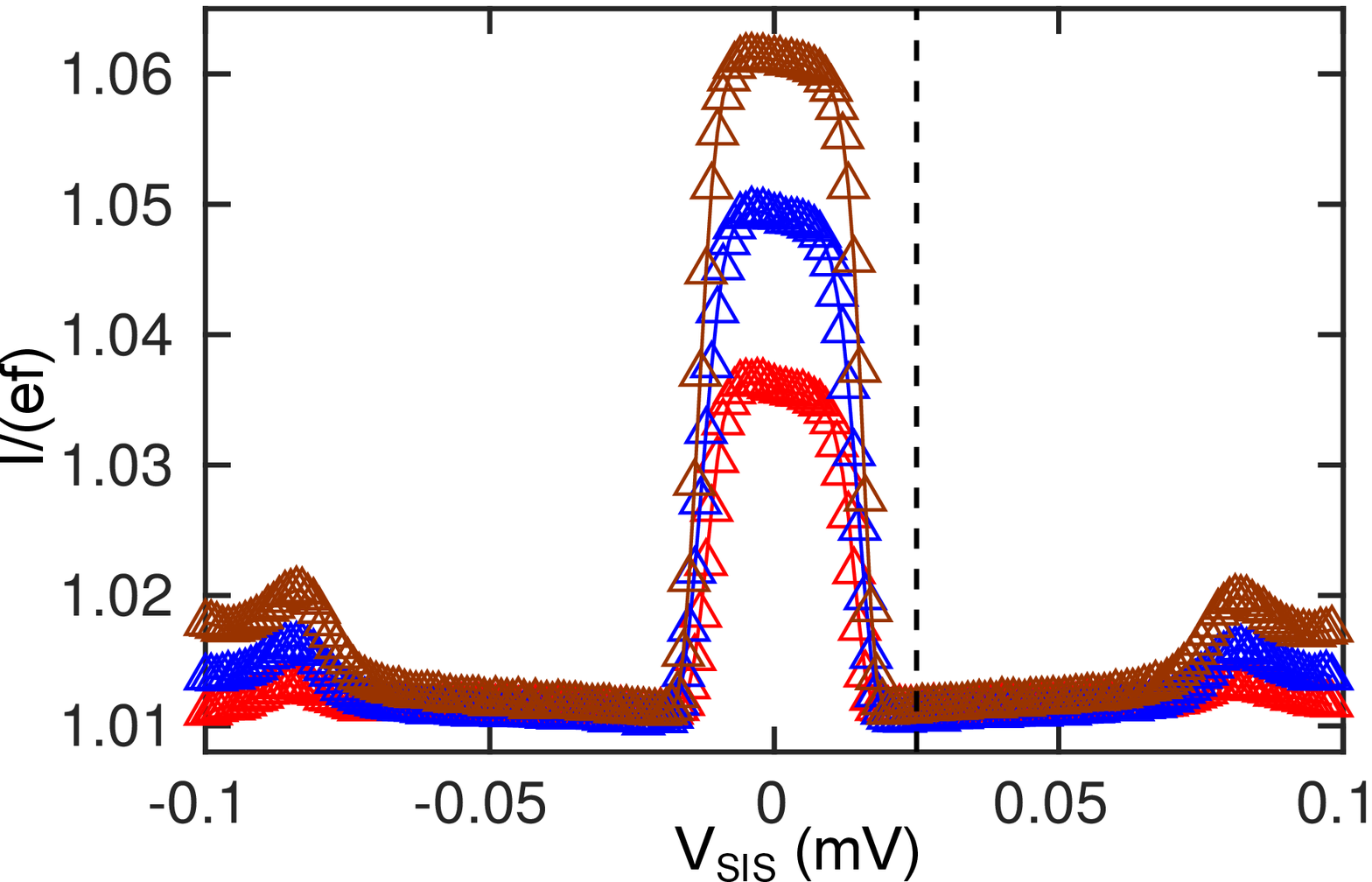}
\caption{Measured current at the plateau as a function of the bias applied to the SQUID at a gate frequency of 40 MHz. Red, blue and brown curves correspond to $V_\mr{b}=100,\,140,\,180\,\mathrm{\mu V}$, respectively.}
\label{f7}
\end{figure}

As expected, the narrow lead cools down when a bias voltage $V_\mathrm{SIS}$ is applied to the SQUID (see Fig.~\ref{f7}) until a certain optimal point is reached. This bias is designated by the dashed black line in Fig.~\ref{f7}. As can be seen, the optimal cooling occurs at $V_\mathrm{SIS}\approx 25\,\mathrm{\mu eV}$, a lower bias than for samples A and B due to the smaller difference $|\Delta_2-\Delta_1|$. Notice that the behaviour shown in Fig.~\ref{f7} is qualitatively the same as that observed in samples A and B.

\begin{figure}[ht!]
\includegraphics[scale=0.75]{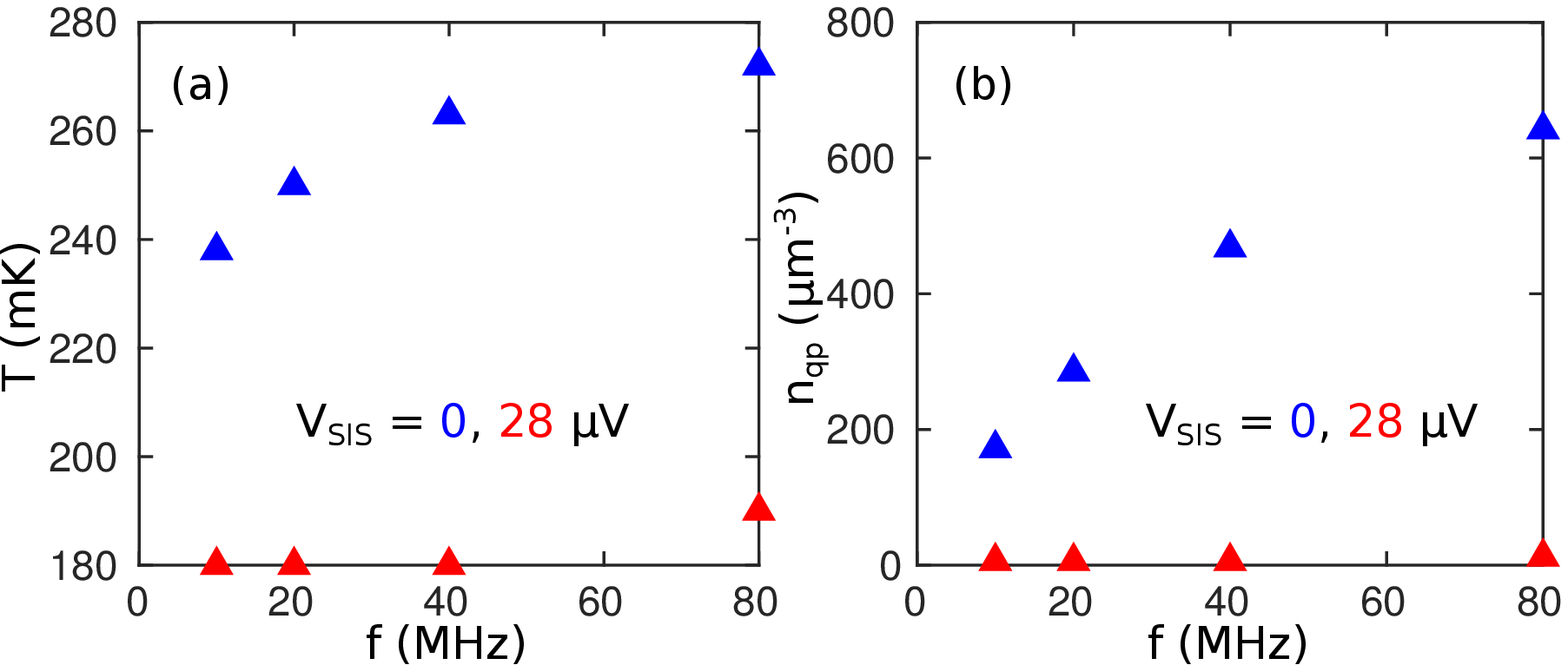}
\caption{(a) Narrow lead temperature as a function of the gate signal frequency. (b) Estimated qp density as function of the gate signal frequency.}
\label{f5}
\end{figure}

In order to estimate the qp reduction in the narrow lead, the device was first operated in the cooler-off regime ($V_\mathrm{SIS}=0\,\mathrm{\mu V}$). Using the same procedure as for samples A and B, the narrow lead temperature and excitation densities were determined in the turnstile operation at different gate signal frequencies by comparing these measurements with simulations. The results can be seen in Fig.~\ref{f5}. It can be seen that the densities are much lower than in the main measurements, which can be understood considering that qp's are more costly to generate due to the higher superconducting gap $\Delta_1$.

In addition, the system was operated near the optimal cooling point ($V_\mathrm{SIS}=28\,\mathrm{\mu V}$). It was observed that the qp density of the narrow lead varies only slightly throughout the different employed frequencies. The temperatures (and excitation densities) remain approximately at $180\,\mathrm{mK}$ ($5.98\,\mathrm{\mu m^{-3}}$) for 10, 20 and 40 MHz, and there is a slight increase to $190\,\mathrm{mK}$ ($12.32\,\mathrm{\mu m^{-3}}$) at $f=80\,\mr{MHz}$. This could be due to an increased recombination rate that is only overcome at 80 MHz. Again, the qp densities have been reduced by an order of magnitude with respect to the cooler-off case.

\subsection{Reference sample without $\mr{S_1IS_2}$ cooler}
\begin{figure}[ht!]
\includegraphics[scale=0.5]{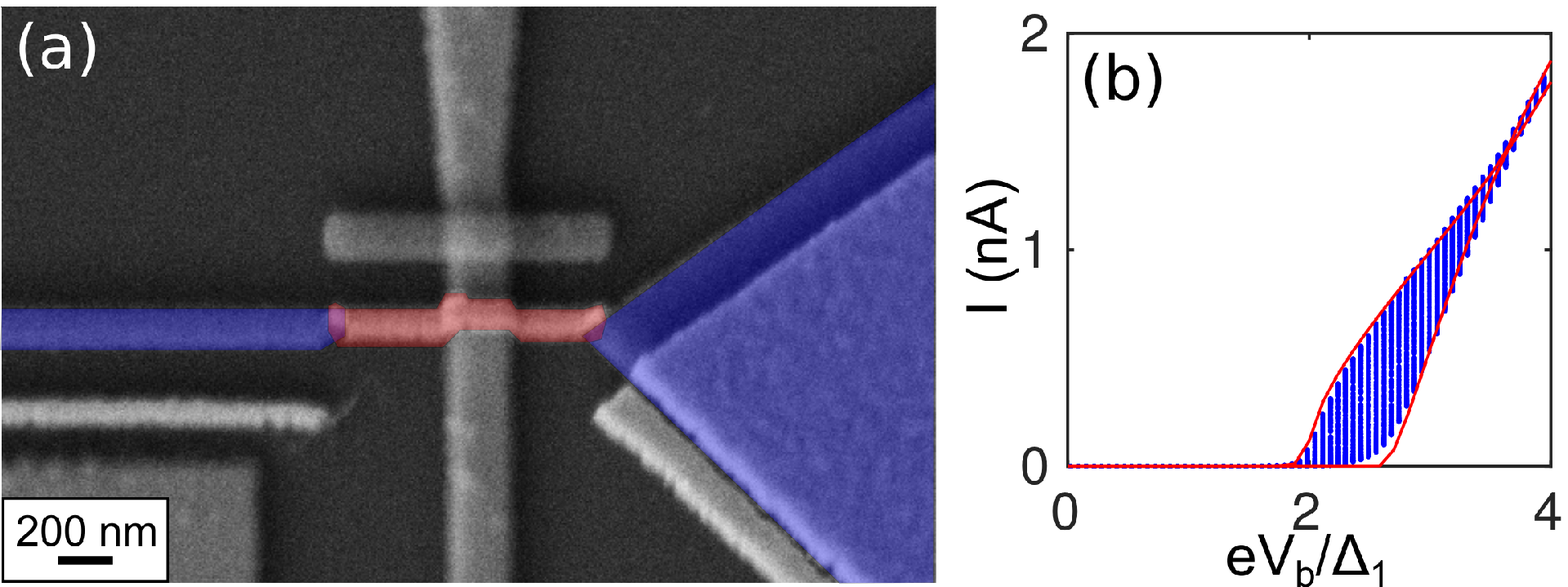}
\caption{Reference sample. (a) Red indicates the normal metal island and blue the superconducting leads. (b) IV curve of the SET in the DC regime. Blue dots are experimental data while the red lines are simulations for extreme values of the gate voltage.}
\label{f1}
\end{figure}

In order to check if the unbiased $\mr{S_1IS_2}$ junctions affect the excitation density a reference sample without the SQUID was measured. As can be seen in Fig.~\ref{f1}(a), the geometry of the reference sample is otherwise the same as for samples A and B described in the main text. After comparing the DC measurements with simulations, it was determined that $\Delta_1=190\;\mu\mathrm{eV}$, $E_\mr{c}=0.63\Delta_1$, $\rt=209.5\,\mathrm{k\Omega}$, $\eta=8\times 10^{-5}$, and $A_\mathrm{ch}=20\,\mathrm{nm^2}$.

\begin{figure}[ht!]
\includegraphics[scale=0.7]{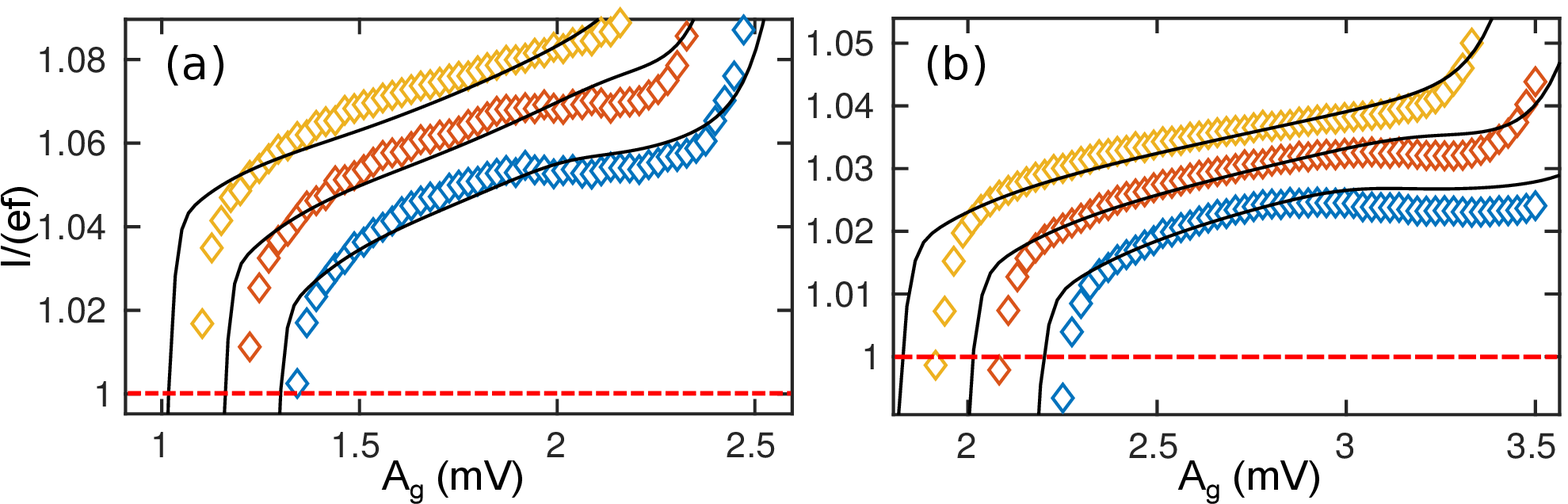}
\caption{Current pumping regime in the reference sample. Open diamonds are measured data and solid black lines are simulated. Blue, red and yellow correspond to bias voltages $V_\mr{b}=100,\, 140\,\mr{and}\, 180\,\mathrm{\mu V}$, respectively. The gate signal has a frequency of (a) 10 MHz and (b) 40 MHz.}
\label{f6}
\end{figure}

\begin{figure}[ht!]
\includegraphics[scale=0.5]{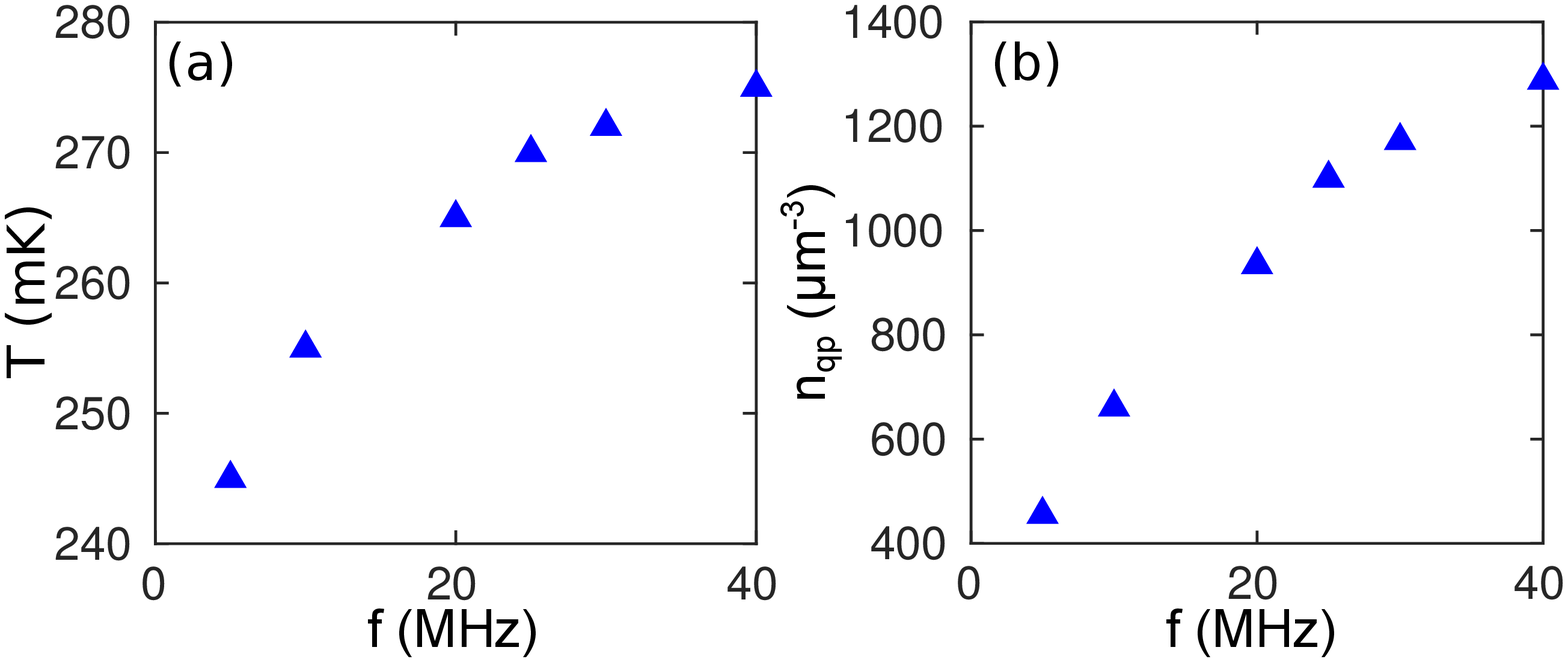}
\caption{Qp density variation as function of drive frequency. (a) Narrow lead temperature as a function of the gate signal frequency for the reference sample. (b) Estimated qp density as function of the gate signal frequency, corresponding to temperatures in panel (a).}
\label{f2}
\end{figure}

Following the same steps as with samples A and B, current measurements in the turnstile operation of the reference SET were done and compared to simulations as shown in Fig.~\ref{f6}. From these comparisons the temperature of the narrow lead is again determined at different driving frequencies (see Fig.~\ref{f2}(a)) and finally the qp density is extracted (see Fig.~\ref{f2}(b)).

It is possible to see in Fig.~\ref{f2}(b) that the values of $n_\mathrm{qp}$ are similar to those of samples A and B in the same driving frequency range and, also the zero frequency limit is in agreement with the main measurements. In general, the $n_\mr{qp}$ values are slightly lower than for samples A and B since in the reference device the superconducting gap $\Delta_1$ is slightly higher. In conclusion, the bare presence of the $\mr{S_1IS_2}$ SQUID does not affect significantly the excitation population in the narrow superconducting lead.

\section{RESISTIVITY MEASUREMENTS}

\begin{figure}[ht!]
\includegraphics[scale=0.9]{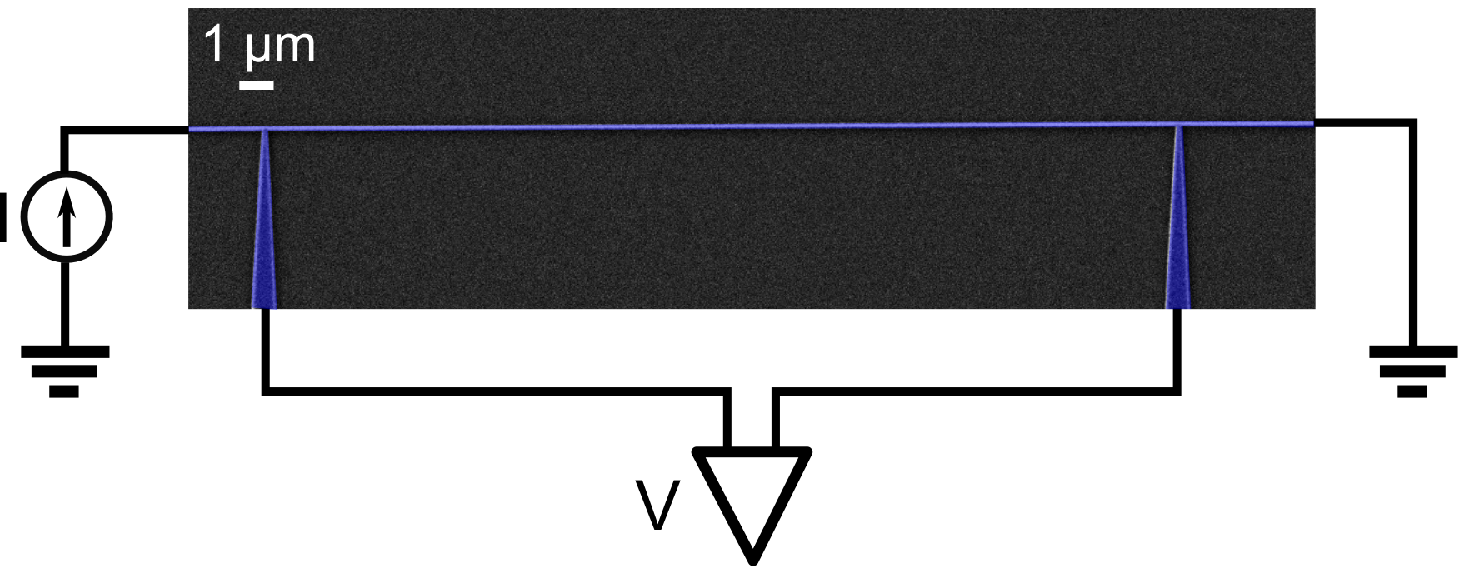}
\caption{A scanning electron micrograph of one of the test wires, used in the the resistivity measurements.}
\label{f10}
\end{figure}

The resistivity of aluminum was measured on separate samples using the four probe method, see Fig.~\ref{f10} for the experimental set up and a typical sample. The test wires were fabricated using a process identical to the full samples. The measurements were done with four different wire geometries, all with a thickness of $70\,\mathrm{nm}$ and lateral dimensions $25\,\mathrm{\mu m}\times 100\,\mathrm{nm}$, $25\,\mathrm{\mu m}\times 200\,\mathrm{nm}$, $50\,\mathrm{\mu m}\times 100\,\mathrm{nm}$, and $50\,\mathrm{\mu m}\times 200\,\mathrm{nm}$, which are resemblant in shape to the long leads of the actual devices. Two batches of test samples were fabricated with differing aluminum deposition rates, $2\,\mathrm{Å/s}$ and $15\,\mathrm{Å/s}$, respectively. The measurements were done at room temperature ($\sim 298\,\mathrm{K}$) and by immersing the sample into liquid nitrogen ($\sim 77\,\mathrm{K}$). The results of these procedures are shown in Table \ref{t1} as the mean value ($\rho$) and the standard deviation ($\sigma$) of a distribution of measurements in units of $\mathrm{\Omega nm}$.

\begin{table}[ht!]
\begin{tabular}{c | c c}
\hline
\hline
Deposition rate $\left(\mr{Å/s}\right)$ & 2 & 15 \\
 \hline
$\rho_{300\,\mr{K}}\,\left(\mathrm{\Omega nm}\right)$ & $51.7$ & $33.1$ \\
$\rho_{77\,\mr{K}}\,\left(\mathrm{\Omega nm}\right)$ & $31.4$ & $13.7$ \\
$\sigma_{300\,\mr{K}}\,\left(\mathrm{\Omega nm}\right)$ & $8.4$ & $3.7$ \\
$\sigma_{77\,\mr{K}}\,\left(\mathrm{\Omega nm}\right)$ & $9.1$ & $2.8$ \\
Number of samples & $20$ & $18$ \\
\hline
\hline
\end{tabular}
\caption{Measured resistivity data from the four probe method in aluminum nanowires. The mean ($\rho$) and the standard deviation ($\sigma$) for a series of measurements done at different conditions (see the text) are shown, together with the number of samples employed. The resistivity of bulk aluminum at room temperature is $27.09\,\mr{\Omega nm}$~\citep{Lide2004}.}
\label{t1}
\end{table}

Since the aluminum deposition rate for the measured SINIS SETs was $\sim 2\,\mathrm{Å/s}$, we estimate that the resistivity of the long leads in the normal state at 77 K is around $31\,\mathrm{\Omega nm}$. Furthermore, immersing a few test wires in liquid helium ($\sim 4.2\,\mr{K}$) allows us to expect a roughly $10\%$ decrease in resistivity ($\rho_{4.2\,\mr{K}}\approx 28\,\mr{\Omega nm}$) compared to the values measured at 77 K.